\documentclass[aps,prb,twocolumn,superscriptaddress,showpacs]{revtex4-1}%%% <===== double-column style

\usepackage{color}
\usepackage{strike}

\usepackage{graphicx}
 
\begin{document}

\title{Second-Order Phase Transition in Heisenberg Model on Triangular Lattice with Competing Interactions}

\author{Ryo Tamura}
\email[]{tamura.ryo@nims.go.jp}
\affiliation{International Center for Young Scientists, National Institute for Materials Science, 1-2-1, Sengen, Tsukuba-shi, Ibaraki, 305-0047, Japan}

\author{Shu Tanaka}
\email[]{shu-t@chem.s.u-tokyo.ac.jp}
\affiliation{Department of Chemistry, University of Tokyo, 7-3-1, Hongo, Bunkyo-ku, Tokyo, 113-0033, Japan}

\author{Naoki Kawashima}
\email[]{kawashima@issp.u-tokyo.ac.jp}
\affiliation{Institute for Solid State Physics, University of Tokyo, 5-1-5, Kashiwanoha, Kashiwa-shi, Chiba, 277-8581, Japan}

\begin{abstract}
We discover an example where the dissociation of the $Z_2$ vortices occurs at the second-order phase transition point.
We investigate the nature of phase transition in a classical Heisenberg model on a distorted triangular lattice with competing interactions.
The order parameter space of the model is SO(3)$\times Z_2$.
The dissociation of the $Z_2$ vortices which comes from SO(3) and a second-order phase transition with $Z_2$ symmetry breaking occur at the same temperature.
We also find that the second-order phase transition belongs to the universality class of the two-dimensional Ising model.
\end{abstract}


%\keywords{}

\maketitle

Geometrically frustrated magnets (GFMs) have been studied exhaustively in a wide range of areas.\cite{Toulouse-1977,Liebmann-1986,Kawamura-1998,Diep-2005}
In GFMs, occurrence of novel phase transitions associated with unconventional order parameters, existence of exotic spin structure, emergence of functional response to external fields, and appearance of unconventional dynamical nature have been predicted theoretically and found experimentally.
A number of examples in which a theoretical calculation gives close agreement with experimental results of GFMs have been reported.\cite{Kageyama-1999,Ramirez-1999,Hertog-2000,Castelnovo-2008,Ishii-2011}
Moreover, since recently there have been found new GFMs which exhibit a theoretically uninvestigated nature,\cite{Nakatsuji-2012,Yoshida-2012} theoretical studies of GFMs and investigations of resultant predictions have increased importance.

In two-dimensional (2D) frustrated models with continuous spins, vortices often appear depending on the symmetry of spin and play an important role in finite-temperature properties.
Although the long-range order of spins is prohibited by the Mermin-Wagner theorem,\cite{Mermin-1966} a topological phase transition driven by the dissociation of vortices can occur.
For example, in the Heisenberg spin systems whose ground state (GS) is a non-collinear structure, $Z_2$ vortices appear due to the order parameter space SO(3).
The dissociation of the $Z_2$ vortices occurs at finite temperature, which is called a $Z_2$ vortex transition.\cite{Kawamura-1984,Okubo-2010}
On the contrary, the order parameter space of the Heisenberg spin systems where the ground state is collinear, such as ferromagnetic state, is $S_2$.
In this case, since there is no topological point defect, no topological phase transition takes place.
In contrast, the Kosterlitz-Thouless transition driven by the dissociation of the $Z$ vortices can occur in the {\it XY} models where the order parameter space is {\it U}(1).

Next let us consider continuous spin systems with competing interactions.
The order parameter space is the direct product between global rotational symmetry of spin and discrete lattice rotational symmetry.
Phase transitions occur corresponding to the order parameter space.\cite{Chandra-1990,Loison-2000,Weber-2003,Tamura-2008,Tamura-2011,Stoudenmire-2009,Okumura-2010,Okubo-2012,Jin-2012}
If the topological point defect exists, both dissociation of vortices and discrete symmetry breaking occur.
If this is the case, two types of behavior can be considered.
In one, phase transitions occur successively.\cite{Loison-2000}
In the other, phase transitions occur simultaneously.
Very recently, some examples of the latter case have been found, in which, the dissociation of vortices occurs at the first-order phase transition point.\cite{Okumura-2010,Tamura-2011}
The origin of the first-order phase transition is breaking of the discrete lattice rotational symmetry.
For example, in Heisenberg spin systems where the order parameter space is SO(3)$\times C_3$, $Z_2$ vortices dissociate at the first-order phase transition point.\cite{Okumura-2010,Tamura-2011}
In addition, in an {\it XY} spin system whose order parameter space is {\it U}(1)$\times Z_6$, the dissociation of $Z$ vortices occurs at the first-order phase transition point.\cite{Tamura-2011}
As described above, a quest for peculiar phase transitions relating to the vortex dissociation has been done for a long time in frustrated spin systems.
However, a second-order phase transition accompanying the dissociation of vortices has never been discovered.

In this paper, we discover an example where the $Z_2$ vortices dissociate at the second-order phase transition point.
We consider a model where the order parameter space is SO(3)$\times Z_2$.
The model under consideration is the Heisenberg model with nearest-neighbor and third nearest-neighbor interactions ($J_1$-$J_3$ model) on a distorted triangular lattice.
We find that the model exhibits a second-order phase transition with the $Z_2$ symmetry breaking and the dissociation of the $Z_2$ vortices at the same temperature.
We also confirm that the second-order phase transition belongs to the universality class of the 2D Ising model.

\begin{figure}[t]
\includegraphics[width=8.4cm]{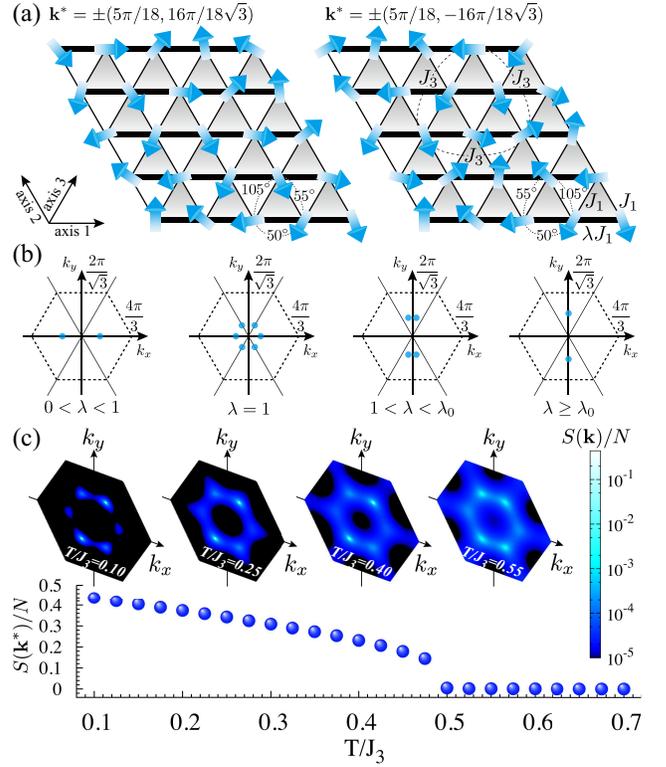}%
\caption{
\label{fig:lattice}
(Color online)
(a) Schematic picture of two distinct GSs of the distorted $J_1$-$J_3$ model for $J_1/J_3=-0.4926\cdots$ and $\lambda=1.308\cdots$. The bold lines indicate $\lambda J_1$. 
(b) $\lambda$-dependence of ${\bf k}^*$ denoted by the circles in the first Brillouin zone depicted by the dotted hexagon.
(c) (upper panel) Structure factors in the first Brillouin zone at several temperatures.
(lower panel) Structure factor $S({\bf k})$ at one of ${\bf k}^*$ as a function of temperature.
The lattice size is $L=144$.
}
\end{figure}

We consider the following Hamiltonian:
\begin{eqnarray}
 \nonumber 
  {\cal H} &&= 
  \lambda J_1 \sum_{\langle i,j \rangle_{\rm axis\, 1}} {\mathbf s}_i \cdot {\mathbf s}_j
  + J_1 \sum_{\langle i,j \rangle_{\rm axis\, 2,3}} {\mathbf s}_i \cdot {\mathbf s}_j\\
 \label{eq:Hamiltonian}
 &&+ J_3 \sum_{\langle\langle i,j \rangle\rangle} {\mathbf s}_i \cdot {\mathbf s}_j,
  \qquad
  \lambda > 0, \, J_3 > 0,
\end{eqnarray}
where the first term represents nearest-neighbor interactions along axis 1, the second term denotes nearest-neighbor interactions along axes 2 and 3, and the summation in the third term takes over the third nearest-neighbor spin pairs [see Fig.~\ref{fig:lattice}(a)].
The variable ${\bf s}_i$ is the three-dimensional vector spin of unit length, and $\lambda$ is a uniaxial distortion parameter.
Let the number of spins be $N(=L\times L)$, where $L$ is the linear dimension.
We hereafter refer to this model as the distorted $J_1$-$J_3$ model.
The model for $\lambda = 1$ is equivalent to the model studied in Ref.~21. 
In this case, a first-order phase transition with breaking of the $C_3$ symmetry occurs when the $C_3$ symmetry is broken in the GS ({\it i.e.} $-4 < J_1/J_3 < 0$).
Since in the GS for the other region of $J_1/J_3$ for $\lambda=1$, no discrete symmetry is broken, we focus on the parameter region in this paper.

In general, the GS of the classical Heisenberg model can be represented by the wave vector ${\bf k}^*$ at which the Fourier transform of interactions $J({\bf k})$ is minimized.
Hereafter, the lattice constant is set to unity.
In the distorted $J_1$-$J_3$ model, $J({\bf k})$ is given by
\begin{eqnarray}
\nonumber
\frac{J({\bf k})}{NJ_3}
&&= \frac{\lambda J_1}{J_3} \cos k_x 
+ \frac{2 J_1}{J_3} \cos \frac{k_x}{2} \cos \frac{\sqrt{3}k_y}{2}\\
\label{eq:gs-jk}
&&+ \cos 2 k_x
+ 2 \cos k_x \cos \sqrt{3}k_y. \label{eq:Jk}
\end{eqnarray}
Notice that the spiral-spin structure represented by ${\bf k}$ and that by $-{\bf k}$ are the same structure in the Heisenberg model.
Figure~\ref{fig:lattice}(b) summarizes positions of ${\bf k}^*$ in the first Brillouin zone depending on $\lambda$.
When $0 < \lambda < 1$, there are two wave vectors ${\bf k}^*=\pm(k_x^*\neq 0,k_y^*=0)$.
In this case, no discrete symmetry is broken in the GS.
For only $\lambda=1$, there are six wave vectors ${\bf k}^*$ and the $C_3$ symmetry is broken in the GS.\cite{Tamura-2011}
If we increase $\lambda$ from unity, there are four wave vectors ${\bf k}^*$.
Then, the $Z_2$ symmetry is broken in the GS.
Moreover, for large $\lambda$ ($> \lambda_0$ obtained by Eq.~(\ref{eq:gs-jk})), the wave vector ${\bf k}^*$ is given by ${\bf k}^*=\pm(k_x^*=0, k_y^*\neq 0)$ and no discrete symmetry is broken in the GS.
In this way, the broken symmetry in the GS can be changed by tuning $\lambda$.
In the following, we focus on the parameter region $1 < \lambda < \lambda_0$, where the order parameter space is SO(3)$\times Z_2$. 
The nature of the phase transition in systems with SO(3)$\times Z_2$ has not yet been investigated.

In order to obtain the equilibrium values of physical quantities with high accuracy, we perform the Monte Carlo simulation using the single-spin-flip heat-bath method and the over-relaxation method.\cite{Creutz-1987,Kanki-2005}
First, we study the nature of the phase transition in the distorted $J_1$-$J_3$ model with the periodic boundary condition.
To avoid extra stress caused by the boundary effect, we use the parameter set such that the GS can be represented by the commensurate wave vector.
We consider the case where the wave vectors ${\bf k}^*$ which minimize $J({\bf k})$ are ${\bf k}^* = \pm(5 \pi/18, 16 \pi/18 \sqrt{3})$ and $\pm(5 \pi/18, -16 \pi/18 \sqrt{3})$.
In the GS spin configuration,
the angle between nearest-neighbor spins along axis 1 is $50$ deg and that along one of the axes 2 and 3 is $55$ deg and the other is $105$ deg, as shown in Fig.~\ref{fig:lattice}(a).
We obtain $J_1/J_3=-0.4926\cdots$ and $\lambda = 1.308\cdots$ using Eq.~(\ref{eq:gs-jk}).
Figure~\ref{fig:lattice}(c) shows temperature $T$ dependence of the structure factor at ${\bf k}^*=(5 \pi/18, 16 \pi/18 \sqrt{3})$ for $L=144$ when the GS spin configuration is represented as the left panel of Fig.~\ref{fig:lattice}(a).
The structure factor is defined as
\begin{eqnarray}
 S({\bf k}) := \frac{1}{N} \sum_{i,j} \langle {\bf s}_i \cdot {\bf s}_j \rangle \, {\rm e}^{-i {\bf k}\cdot({\bf r}_i - {\bf r}_j)},
\end{eqnarray}
where the summation takes over all of the spin pairs and ${\bf r}_i$ represents the position vector of the $i$-th site.
Throughout the paper, $\langle {\cal O} \rangle$ denotes the equilibrium value of the physical quantity ${\cal O}$ and the Boltzmann constant is set to unity.
As temperature decreases, $S({\bf k}^*)$ monotonically increases.
The structure factor in the first Brillouin zone at several temperatures is also shown in Fig.~\ref{fig:lattice}(c), which suggests that the $Z_2$ symmetry is broken at low temperature.\cite{Footnote}

\begin{figure}[t]
\includegraphics[width=8.5cm]{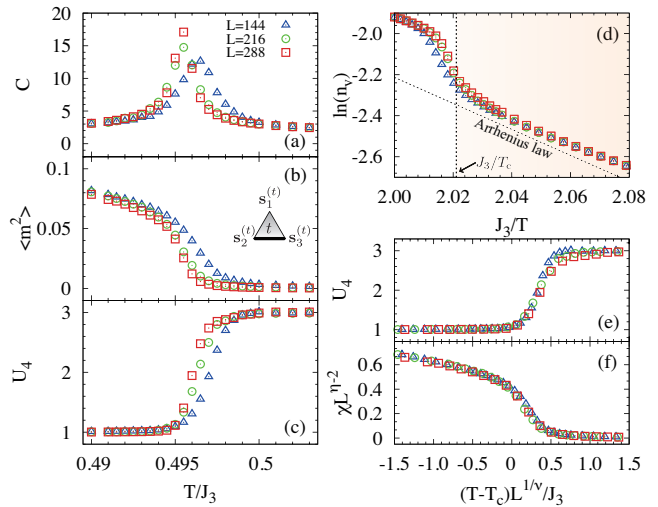}
\caption{
\label{graph:pqs-pbc}
(Color online) Temperature dependence of equilibrium physical quantities of the distorted $J_1$-$J_3$ model for $J_1/J_3=-0.4926\cdots$ and $\lambda =1.308\cdots$.
(a) Specific heat $C$.
(b) Square of the order parameter $\langle m^2 \rangle$.
(c) Binder ratio $U_4$.
(d) Log of number density of $Z_2$ vortex $n_{\rm v}$ versus $J_3/T$.
The dotted vertical line indicates the transition temperature $T_{\rm c}/J_3=0.4950(5)$.
(e) and (f) Finite-size scaling of the Binder ratio $U_4$ and that of the susceptibility $\chi$ using the critical exponents of the 2D Ising model ($\nu=1$ and $\eta=1/4)$ and the transition temperature.
Error bars are omitted for clarity since their sizes are smaller than the symbol sizes.
}
\end{figure}

We calculate the temperature dependence of specific heat for $L=144-288$.
The specific heat is calculated as
\begin{eqnarray}
 C = (\langle E^2 \rangle - \langle E \rangle^2)/T^2,
\end{eqnarray}
where $E$ is the internal energy per site.
Figure~\ref{graph:pqs-pbc}(a) shows the specific heat as a function of temperature and indicates the existence of a phase transition at finite temperature.
To confirm the occurrence of a spontaneous $Z_2$ symmetry breaking, we define an order parameter $m$ by using a local quantity $\kappa^{(t)}$ at each upward elementary triangle $t$ depicted by the shaded triangles in Fig.~\ref{fig:lattice}(a):
\begin{eqnarray}
 \kappa^{(t)} := {\mathbf s}_1^{(t)} \cdot ({\mathbf s}_2^{(t)} - {\mathbf s}_3^{(t)}),
  \qquad
  m := \sum_t \kappa^{(t)} / N,
\end{eqnarray}
where the definition of the subscript of spins is shown in Fig.~\ref{graph:pqs-pbc}(b) and the summation takes over all upward elementary triangles.
Since $m$ represents the difference between the relative angle of nearest-neighbor spins along axis 2 and that along axis 3, it appropriately describes the $Z_2$ symmetry breaking in our model.
Figures \ref{graph:pqs-pbc}(b) and \ref{graph:pqs-pbc}(c) show the temperature dependence of $\langle m^2 \rangle$ and $U_4 := \langle m^4 \rangle/\langle m^2 \rangle^2$, respectively.
These figures indicate the existence of the second-order phase transition with breaking of the $Z_2$ symmetry at $T_{\rm c}/J_3 = 0.4950(5)$, where the Binder ratios for some lattice sizes cross.

In antiferromagnetic Heisenberg models on a triangular lattice, 
the dissociation of the $Z_2$ vortices occurs at finite temperature.\cite{Kawamura-1984,Kawamura-2010}
In order to confirm the dissociation of the $Z_2$ vortices in our model,
we calculate the number density of the $Z_2$ vortices $n_{\rm v}$ by using the same manner as in Ref.~13.
A plot of $\ln n_{\rm v}$ versus $J_3/T$ in our model is shown in Fig.~\ref{graph:pqs-pbc}(d),
and it is confirmed that $n_{\rm v}$ obeys well the Arrhenius law below $T_{\rm c}$.
This result indicates that the dissociation of the $Z_2$ vortices occurs at the second-order phase transition point.

To clarify the universality class of the phase transition, we perform the finite-size scaling using the following relations:
\begin{eqnarray}
 U_4 \propto f[(T-T_{\rm c}) L^{1/\nu}],
  \,
  \chi \propto L^{2-\eta} g[(T-T_{\rm c}) L^{1/\nu}],
\end{eqnarray}
where the susceptibility $\chi$ is defined as $\chi := N J_3 \langle m^2 \rangle/T$ and $f(\cdot)$ and $g(\cdot)$ are scaling functions.
The finite-size scaling results using $\nu=1$ and $\eta=1/4$ which are the critical exponents of the 2D Ising model and the obtained $T_{\rm c}$ are shown in Figs.~\ref{graph:pqs-pbc}(e) and \ref{graph:pqs-pbc}(f).
Since all the data collapse onto scaling functions,
it is confirmed that the second-order phase transition in our model belongs to the universality class of the Ising model.

\begin{figure}[t]
\includegraphics[width=8.4cm]{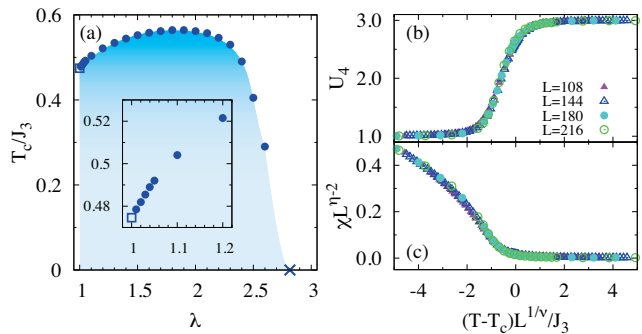}
\caption{
\label{graph:obc}
(Color online) (a) Phase diagram of the distorted $J_1$-$J_3$ model for $J_1/J_3=-0.7342\cdots$.
The inset is an enlarged view.
The open square indicates the transition temperature for $\lambda=1$ where a first-order phase transition with $C_3$ symmetry breaking occurs.\cite{Tamura-2011}
The solid circles represent transition temperatures at which a second-order phase transition with $Z_2$ symmetry breaking occurs.
(b) and (c) Finite-size scaling of the Binder ratio $U_4$ and that of the susceptibility $\chi$ for $\lambda = 1.5$ using $\nu=1$ and $\eta=1/4$ which are the critical exponents of the 2D Ising model.
Error bars are omitted for clarity since their sizes are smaller than the symbol sizes.
}
\end{figure}

Next, to obtain the relationship between $\lambda$ and $T_{\rm c}$,
we consider the case of $J_1/J_3=-0.7342\cdots$ which was used in Ref.~21 by changing the value of $\lambda$.
For $\lambda=1$, the model exhibits a first-order phase transition with breaking of the $C_3$ symmetry at $T_{\rm c}/J_3=0.4746(1).$\cite{Tamura-2011}
Here we study the nature of the phase transition of the distorted $J_1$-$J_3$ model with the open boundary condition.
From the analysis of the GS as explained before, a phase transition with breaking of the $Z_2$ symmetry is expected to take place for $1 < \lambda < \lambda_0(=2.8155\cdots)$ in this case.
By analyzing the Binder ratio, we obtain $\lambda$ dependence of transition temperatures as depicted in Fig.~\ref{graph:obc}(a).
An enlarged view near $\lambda = 1$ is shown in the inset of Fig.~\ref{graph:obc}(a).
This figure indicates that the transition temperature near $\lambda = 1$ smoothly connects to the transition temperature for $\lambda = 1$ and the transition temperature goes continuously to zero when $\lambda \to \lambda_0$.
Figures \ref{graph:obc}(b) and \ref{graph:obc}(c) represent the finite-size scaling of the Binder ratio and that of the susceptibility for $\lambda = 1.5$ using $\nu=1$, $\eta =1/4$, and $T_{\rm c}/J_3=0.5521(1)$ as well as the previous case.
In this case, all the data collapse onto scaling functions.
Thus, we conclude that a second-order phase transition with breaking of the $Z_2$ symmetry occurs and it belongs to the 2D Ising model universality class within calculated $\lambda$.
However, at very close to $\lambda = 1$, the possibility that first-order phase transition occurs with breaking of the $Z_2$ symmetry cannot be denied.
Unexpected phase transition from only underlying symmetry can occur in some cases.\cite{Loison-2000,Tamura-2010,Jin-2012}
If a first-order phase transition with breaking of the $Z_2$ symmetry occurs, a tricritical point should exist and to study its properties such as universality class will be an important topic.
From our observation, it is difficult to obtain the nature of the phase transition near $\lambda=1$ since the size dependence of physical quantities are significant and we should calculate very large systems with high accuracy.

In this paper, we discovered an example where the second-order phase transition occurs accompanying the $Z_2$ vortex dissociation at finite temperature.
The model under consideration is the classical Heisenberg model on a triangular lattice with three types of interactions: (i) the uniaxially distorted nearest-neighbor ferromagnetic interaction along axis 1 ($\lambda J_1$), (ii) the nearest-neighbor ferromagnetic interaction along axes 2 and 3 ($J_1$), and (iii) the third nearest-neighbor antiferromagnetic interaction ($J_3$).
In this model for $1 < \lambda < \lambda_0$, the order parameter space is SO(3)$\times Z_2$. 
We found the second-order phase transition with spontaneous breaking of the $Z_2$ symmetry in the region. 
The dissociation of the $Z_2$ vortices also occurs at the same temperature.
The dissociation of vortices at the first-order phase transition point has been found in some cases.\cite{Okumura-2010,Tamura-2011}
Then, the relation between the dissociation of vortices and phase transition due to discrete symmetry cannot be complete without our study.
Furthermore, we confirmed that the universality class of the second-order phase transition with the $Z_2$ symmetry breaking is the same as that of the 2D Ising model, despite that the $Z_2$ vortex dissociation occurs at the critical point.
This indicates that the $Z_2$ vortex dissociation never affects the critical phenomena in our model.

Let us next mention the relations to experiments.
In this paper, we have shown that our model exhibits both a first-order phase transition and a second-order phase transition by tuning the uniaxial distortion parameter $\lambda$.
The uniaxial distortion parameter $\lambda$ can be experimentally changed by pressure or/and chemical substitution. 
In some triangular antiferromagnets, uniaxial lattice distortion occurs at a temperature which is significantly higher than that of magnetic ordering.\cite{Kenzelman-2007}
Similar materials which exhibit the obtained results will be synthesized.
Furthermore, it is interesting to consider the nature of the phase transition of stacked triangular lattice systems with competing interactions.
If the order parameter space is SO(3)$\times Z_2$, a long-range order of spins should occur in addition to a phase transition relating to the $Z_2$ symmetry breaking.
It is possible that a novel universality class appears in the model.
Moreover, by applying a magnetic field, a rich phase diagram and novel spin structures in excited states are expected, as well as Ref.~22.

Finally, we emphasize that the obtained results are not restricted to our model and typically occur in the systems having the order parameter space SO(3)$\times Z_2$.
The order parameter space SO(3)$\times Z_2$ appears in the magnetic systems having following conditions:
(i) the spin is Heisenberg type, (ii) the ground state is a spiral spin structure due to competing interactions, and (iii) there is a reflection symmetry of lattice.
Since all of these are mild conditions,
the obtained nature should occur in many frustrated magnetic systems.
In addition, not only in magnetic systems, the obtained nature can also occur in systems having both continuous degrees of freedom belonging to SO(3) symmetry and discrete degrees of freedom described by $Z_2$ symmetry.
For example, the obtained nature might be observed in the systems where the underlying symmetry can be easily changed, such as liquid crystal and spinor BEC.\cite{Sadler-2006,Alexande-2012}
Therefore, the vortex dissociation at a critical point should be ubiquitously observed in 2D systems with two or more degrees of freedom when the order parameter space is SO(3)$\times Z_2$.

We thank Michikazu Kobayashi for useful comments and discussions.
The authors are partially supported by National Institute for Materials Science (NIMS), Grand-in-Aid for Scientific Research (C) (25420698), JSPS Fellows (23-7601), Scientific Research (B) (22340111), and the Computational Materials Science Initiative (CMSI). 
Numerical calculations were performed on supercomputers at the Institute for Solid State Physics, University of Tokyo.


\begin{thebibliography}{99}
\bibitem{Toulouse-1977} % check
G.~Toulouse,
Commun. Phys. (London) {\bf 2}, 115 (1977).

\bibitem{Liebmann-1986} % check
R.~Liebmann,
{\it Statistical Mechanics of Periodic Frustrated Ising Systems}
(Springer-Verlag, Berlin/Heidelberg GmbH, Heidelberg, 1986).

\bibitem{Kawamura-1998} % check
H.~Kawamura,
J. Phys.: Condens. Matter {\bf 10}, 4707 (1998).

\bibitem{Diep-2005} % check
{\it Frustrated Spin Systems},
edited by H.~T.~Diep (World Scientific, Singapore, 2005).

\bibitem{Kageyama-1999} % check
H.~Kageyama, K.~Yoshimura, R.~Stern, N.~V.~Mushnikov, K.~Onizuka, M.~Kato, K.~Kosuge, C.~P.~Slichter, T.~Goto, and Y.~Ueda,
Phys. Rev. Lett. {\bf 82}, 3168 (1999).

\bibitem{Ramirez-1999} % check
A.~P.~Ramirez, A.~Hayashi, R.~J.~Cava, R.~Siddharthan, and B.~S.~Shastry,
Nature (London) {\bf 399}, 333 (1999).

\bibitem{Hertog-2000} % check
B.~C.~den Hertog and M.~J.~P.~Gingras,
Phys. Rev. Lett. {\bf 84}, 3430 (2000).

\bibitem{Castelnovo-2008} % check
C.~Castelnovo, R.~Moessner, and S.~L.~Sondhi,
Nature (London) {\bf 451}, 42 (2008).

\bibitem{Ishii-2011} % check
R.~Ishii, S.~Tanaka, K.~Onuma, Y.~Nambu, M.~Tokunaga, T.~Sakakibara, N.~Kawashima, Y.~Maeno, C.~Broholm, D.~P.~Gautreaux, J.~Y.~Chan, and S.~Nakatsuji,
Europhys. Lett. {\bf 94}, 17001 (2011).

\bibitem{Yoshida-2012} % check
H.~Yoshida, J.~Yamaura, M.~Isobe, Y.~Okamoto, G.~J.~Nilsen, and Z.~Hiroi,
Nat. Commun. {\bf 3}, 860 (2012).

\bibitem{Nakatsuji-2012} % check
S.~Nakatsuji, K.~Kuga, K.~Kimura, R.~Satake, N.~Katayama, E.~Nishibori, H.~Sawa, R.~Ishii, M.~Hagiwara, F.~Bridges, T.~U.~Ito, W.~Higemoto, Y.~Karaki, M.~Halim, A.~A.~Nugroho, J.~A.~Rodriguez-Rivera, M.~A.~Green, and C.~Broholm,
Science {\bf 336}, 559 (2012).

\bibitem{Mermin-1966} % check
N.~D.~Mermin and H.~Wagner,
Phys. Rev. Lett. {\bf 17}, 1133 (1966).

%\bibitem{Miyashita-1984} % check
%S.~Miyashita and H.~Shiba,
%J. Phys. Soc. Jpn. {\bf 53}, 1145 (1984).

\bibitem{Kawamura-1984} % check
H.~Kawamura and S.~Miyashita,
J. Phys. Soc. Jpn. {\bf 53}, 4138 (1984).

\bibitem{Okubo-2010} % check
T.~Okubo and H.~Kawamura,
J. Phys. Soc. Jpn. {\bf 79}, 084706 (2010).

\bibitem{Chandra-1990} % check
P.~Chandra, P.~Coleman, and A.~I.~Larkin,
Phys. Rev. Lett. {\bf 64}, 88 (1990).

\bibitem{Loison-2000} % check
D.~Loison and P.~Simon,
Phys. Rev. B {\bf 61}, 6114 (2000).

\bibitem{Weber-2003} % check
C.~Weber, L.~Capriotti, G.~Misguich, F.~Becca, M.~Elhajal, and F.~Mila,
Phys. Rev. Lett. {\bf 91}, 177202 (2003).

\bibitem{Tamura-2008} % check
R.~Tamura and N.~Kawashima,
J. Phys. Soc. Jpn. {\bf 77}, 103002 (2008).

\bibitem{Stoudenmire-2009} % check
E.~M.~Stoudenmire, S.~Trebst, and L.~Balents,
Phys. Rev. B {\bf 79}, 214436 (2009).

\bibitem{Okumura-2010} % check
S.~Okumura, H.~Kawamura, T.~Okubo, and Y.~Motome,
J. Phys. Soc. Jpn. {\bf 79}, 114705 (2010).

\bibitem{Tamura-2011} % check
R.~Tamura and N.~Kawashima,
J. Phys. Soc. Jpn. {\bf 80}, 074008 (2011).

\bibitem{Okubo-2012} % check
T.~Okubo, S.~Chung, and H.~Kawamura,
Phys. Rev. Lett. {\bf 108}, 017206 (2012).

\bibitem{Jin-2012} % check
S.~Jin, A.~Sen, and A.~W.~Sandvik,
Phys. Rev. Lett. {\bf 108}, 045702 (2012).

\bibitem{Creutz-1987} %%% over relaxation   % check
M.~Creutz,
Phys. Rev. D {\bf 36}, 515 (1987).

\bibitem{Kanki-2005} %%% over relaxation   % check
K.~Kanki, D.~Loison, and K.~D.~Schotte,
Eur. Phys. J. B {\bf 44}, 309 (2005).

\bibitem{Footnote}
Note that, since the long-range order of spins at finite-temperature is prohibited by the Mermin-Wagner theorem in our model, $S({\bf k})$ for arbitrary ${\bf k}$ should be vanished in the thermodynamic limit. 
Then, $S({\bf k})$ as shown in Fig.~~\ref{fig:lattice}(c) is just an indication for finite-size systems. 
In order to study the phase transition with discrete symmetry breaking at finite temperature in our model, we should define an appropiate order parameter instead of $S({\bf k})$ and investigate its thermodynamic properties.

\bibitem{Kawamura-2010} % check
H.~Kawamura, A.~Yamamoto, and T.~Okubo,
J. Phys. Soc. Jpn. {\bf 79}, 023701 (2010).

\bibitem{Tamura-2010}
R.~Tamura, S.~Tanaka, and N.~Kawashima,
Prog. Theor. Phys. {\bf 124}, 381 (2010).

\bibitem{Kenzelman-2007}
M.~Kenzelmann, G.~Lawes, A.~B.~Harris, G.~Gasparovic, C.~Broholm, A.~P.~Ramirez, G.~A.~Jorge, M.~Jaime, S.~Park, Q.~Huang, A.~Ya.~Shapiro, and L.~A.~Demianets, 
Phys. Rev. Lett. {\bf 98}, 267205 (2007).

\bibitem{Sadler-2006}
L.~E.~Sadler, J.~M.~Higbie, S.~R.~Leslie, M.~Vengalattore, and D.~M.~Stamper-Kurn,
Nature (London) {\bf 443}, 312 (2006).

\bibitem{Alexande-2012}
G.~P.~Alexander, B.~G.~Chen, E.~A.~Matsumoto, and R.~D.~Kamien,
Rev. Mod. Phys. {\bf 84}, 497 (2012).


\end{thebibliography}
\end{document}